\documentclass[12pt]{article}

\usepackage{braket}
\usepackage{mathrsfs}
 \usepackage{extpfeil}
 \usepackage{graphics}
\usepackage{epstopdf}
 \usepackage{cite}
\numberwithin{equation}{section}
\begin{document}

\begin{center}
{\LARGE{\bf{     On the Derivation and Interpretation of the Poincar\'e-Maxwell Group      }}}
\end{center}

\bigskip\bigskip

\begin{center}
 Przemys\l aw Brzykcy\footnote{E-mail address:  800289@edu.p.lodz.pl}  
 \end{center}

\begin{center}

{\sl  Institute of Physics, Lodz University of Technology,\\ W\'{o}lcza\'{n}ska 219, 90-924 \L\'{o}d\'{z}, Poland.}\\
\medskip

\end{center}

\vskip 1.5cm
\centerline{\today}
\vskip 1.5cm

\begin{abstract}
The Lie algebra of the Poincar\'e-Maxwell group is derived in a manner that provides the 
interpretation of the equations of motion. It is clarified that the dynamics obtained 
from the orbit method is exactly equivalent to the classical description of the charged particle 
moving in the constant electromagnetic field. The multiplicity of the coadjoint 
orbits of the group under consideration is discussed.
\end{abstract}


 \section{Introduction} \label{sec1} 

In the past years much attention has been attracted by $(2+1)$ dimensional  models 
displaying Galilean symmetry
that include non-commutative geometry \cite{lukierski1997galilean,Duval2000284,Jackiw2000237,Ghosh2006350}. 
In many works devoted to this subject
\cite{0305-4470-34-47-314,Duval:2001hu,1126-6708-2002-06-033,delOlmo20062830,SIGMA}
the Galilei group was studied both in free case and with external fields, also 
its central extensions have been investigated  \cite{1995q.alg.....8020B}.
Other works devoted to the description of a charged particle 
moving in the electromagnetic field include 
\cite{doi:10.1063/1.528891,doi:10.1063/1.528984,PROP:PROP2190380903}.
There has also been some interest in similar considerations in the realm of
relativistic theory \cite{doi:10.1063/1.528891,anyons,PhysRevD.43.1933}
to which subject this paper aspires to contribute to.
Similar techniques has been 
used  in \cite{NH1, Alvarez20071556 , NH2, Zhang ,NH3,NH4,NH5,NH6} to investigate models with Newton-Hooke 
like symmetries in various contexts.

The present work is greatly  motivated by \cite{anyons} and inspired by 
\cite{stichel}.
Its main purpose is to provide the interpretation of the equations of 
motion previously obtained in \cite{anyons} therefore it is a straight 
continuation of the work therein. Recently, it was highlihted \cite{brzykcy} that within the 
framework of the orbit method a physical interpretation is carried by a specific 
realisation of a Lie algebra via smooth functions rather than by an abstract Lie 
algebra.
Towards this goal, a way of deriving 
the Poincar\'e-Maxwell group sparked by \cite{stichel} is devised.
This result displays the link between the approach consisting in     
building the dynamics on the coadjoint orbit of a proper Lie group with the more 
conventional framework of classical mechanics. This clarification on the meaning 
of the equations of motion obtained by the orbit method clears the way towards 
further investigations.

This paper is structured as follows.  Section \ref{PM} provides a derivation of 
the Poincar\'e-Maxwell group in $(2+1)$ dimensions ($PM(2+1)$) carried out 
in a manner which lights up the road towards the interpretation of the equations of motion.  
Section \ref{coorbit} focusses on the different kinds of coadjoint orbits of 
$PM(2+1)$. Finally, in Section \ref{symplectic} a symplectic structure on the 
coadjoint orbit is given and the equations of motion are derived. Also, results 
of Section \ref{PM} are utilised to  produce a clear interpretation of the 
dynamics.
The article closes with some conclusions and comments on the relation with 
similar works in Section \ref{conclusions} where, also some future outlooks are discussed.
 
 \section{The Poincar\'e-Maxwell group}\label{PM} 
 
This section  is focused on  a relativistic particle endowed with a charge $q$ moving in 
 the $xy$ plane under the influence of a constant electric field $\vec{E}=(E_x,E_y,0)$ and 
 a constant magnetic field  perpendicular to the $xy$ plane $\vec{B}=(0,0,B)$. 
 The Lie algebra of the Poincar\'e--Maxwell (PM(2+1)) group will be derived.

 To focus the attention choose the symmetric gauge for the vector potential 
 $
\vec{A} =  \frac{B}{2} [-y,x,0]
$
so that  $
\vec{B} 
     =[ 0,0,B ]
$
and the scalar potential 
$
  \phi= -(E_x\,x  +E_y \,y    )
$
yielding  
$
\vec{E} =-\nabla \phi - \frac{\partial \vec{A}}{\partial t}=[  E_x,E_y,0].
$
The Lagrangian for this system can be written as
\begin{equation}
  \mathcal{L}= -m_0c^2\sqrt{1-\frac{\dot{x}^2  + \dot{y}^2}{c^2}}-\frac{qBy\dot{x}}{2}+\frac{qBx\dot{y}}{2}
  +qE_x x+qE_y y
\end{equation}
which yields  the following generalised momenta
\begin{equation}
  P_x=\frac{m_0 \dot{x}}{\sqrt{1    -\frac{   \dot{x}^2+\dot{y}^2  }{c^2}     
  }}-\frac{qBy}{2},\quad
  P_y=\frac{m_0 \dot{y}}{\sqrt{1    -\frac{   \dot{x}^2+\dot{y}^2  }{c^2}     }}+\frac{qBx}{2}
\end{equation}
and  the Hamiltonian   $ \mathcal{H}=\vec{P} \cdot \dot{\vec{r}}-\mathcal{L} $ 
reads
\begin{equation}
 \mathcal{H} =c \sqrt{
 \left( P_x+\frac{qBy}{2}\right) ^2+
 \left( P_y-\frac{qBx}{2}\right) ^2+m_0^2c^2
 }
 -qE_x x
 -qE_y y.  \label{PM_hamiltonian1}
 \end{equation}
Note that  the coordinates $(x,y,P_x,P_y)$ are canonical by construction.  
 Moreover,  the fields $E_x$, $E_y$ and $B$ will be treated as 
 additional coordinates (keeping in mind that they are constant).
 In order to  fit them into the Hamiltonian framework, one has   to introduce their 
 canonically conjugated momenta denoted $\pi_x$, $\pi_y$ and $\beta$ 
 respectively.
 That is to   assume the Poisson  bracket in the following form
     \begin{equation}\label{PM_poisson_1}
  \begin{split}
      \left\{ F,G \right\} &=  
   \left(
     \frac{\partial F}{\partial x}  \frac{\partial G}{\partial P_x   }
  - \frac{\partial F}{\partial P_x}  \frac{\partial G}{\partial x   }
  +\frac{\partial F}{\partial y}  \frac{\partial G}{\partial P_y   }
  -\frac{\partial F}{\partial P_y}  \frac{\partial G}{\partial y   } 
  \right) \\
     &+ 
    \left(
   \frac{\partial F}{\partial E_x}  \frac{\partial G}{\partial \pi_x   }
  - \frac{\partial F}{\partial \pi_x}  \frac{\partial G}{\partial E_x   }
  +\frac{\partial F}{\partial E_y}  \frac{\partial G}{\partial \pi_y   }
  -\frac{\partial F}{\partial \pi_y}  \frac{\partial G}{\partial E_y   } 
  \right)\\
     &+ 
    \left(
   \frac{\partial F}{\partial B}  \frac{\partial G}{\partial \beta   }
  - \frac{\partial F}{\partial \beta}  \frac{\partial G}{\partial B   }
  \right).
  \end{split}
\end{equation} 
  Note that (\ref{PM_poisson_1}) implies that units of $\pi_i$ and $\beta$ are $\left[ \pi_i \right] = \left[ \mathrm A\, m \,  s^2\right]$ and $\left[\beta\right]  = \left[ \mathrm A\, m^2 \,  s 
  \right]$.
Next step is to find the integrals of motion i.e. the solutions the 
following differential equation
\begin{equation}
  \{ f , \mathcal{H}\} =0 \label{PM_inv_eq1}
\end{equation}
where $f=f(x,y,P_x,P_y,E_x,E_y,\pi_x,\pi_y,B,\beta)$. One obvious solution is $\mathcal{H}$ furthermore, 
since $\mathcal{H}$ does not depend  
on $\pi_x$, $\pi_y$ nor $\beta$ there are solutions  proportional to $E_x$, $E_y$ and $B$. Take them to be
$      \mathcal{B} = qB $,
$      \mathcal{E}_x = -qE_x$ and
$       \mathcal{E}_y = -qE_y$.
One more  solution to (\ref{PM_inv_eq1}) is 
$     \mathcal{J} =  x P_y - yP_x  +E_x \pi_y-E_y \pi_x$.
  
  The functions $\mathcal{H}$, $\mathcal{J}$, $\mathcal{E}_x$, $\mathcal{E}_y$ 
  and $\mathcal{B}$ together with a Poisson bracket (\ref{PM_poisson_1}) form a Lie algebra 
  of a symmetry group of this   system.
 Next, let us cast about the constants of motion i.e. the functions $f=f(t,x,y,P_x,P_y,E_x,E_y,\pi_x,\pi_y,B,\beta)$
  such that
  \begin{equation}
  \{ f , \mathcal{H}\} + \frac{\partial f}{\partial t}=0. \label{PM_inv_eq2}
\end{equation}
 The first two solutions to (\ref{PM_inv_eq2}) have the dimension $\left[\frac{\mathrm{ kg 
 \,m}}{\mathrm{s}}\right]$, just as momentum does and they are of the form
      \begin{equation}
    \begin{split}
      \mathcal{P}_x &=  P_x - \frac{qBy}{2}-  qE_x t,\\
      \mathcal{P}_y &= P_y + \frac{qBx}{2}-  qE_y t,
      \end{split}\label{PM_Pi}
  \end{equation}   
  another two solutions have dimension $\left[  \mathrm {J m}\right]$ and they read 
     \begin{equation}
    \begin{split}   
      \mathcal{K}_x &=  
      x  c \sqrt{{{\left( P_x+\frac{qBy}{2}\right) }^{2}}+{{\left( P_y- \frac{qBx}{2}\right) }^{2}}+m^2c^2} \\ 
      &- c^2 t  \left( P_x -\frac{qBy}{2} +\frac{qE_xt}{2}\right) 
      -\frac{qE_y x y}{2}
      -\frac{ qE_x x^2}{2}  + c^2 B\pi_y  -  E_y \beta,\\
         \mathcal{K}_y &= 
      y c \sqrt{{{\left( P_x+\frac{qBy}{2}\right) }^{2}}+{{\left( P_y- \frac{qBx}{2}\right) }^{2}}+m^2c^2} \\ 
      &- c^2 t \left( P_y  + \frac{qBx}{2} - \frac{qE_yt}{2}\right)  
      -\frac{qE_x x y}{2}
      -\frac{ qE_y y^2}{2}   + c^2 B\pi_x +  E_x \beta.  
       \end{split} \label{PM_Ki}
  \end{equation}   
Consider a set of functions $( \mathcal{B}, \mathcal{E}_x,\mathcal{E}_y, \mathcal{H},    \mathcal{P}_x, \mathcal{P}_y, \mathcal{K}_x, \mathcal{K}_y, \mathcal{J})$
  (they shall be kept  in this order).
  After calculating the Poisson bracket (\ref{PM_poisson_1}) for all the pairs chosen from 
  the aforementioned set 
   $$
    \bordermatrix{
 ~ & \mathcal{B} &\mathcal{E}_x &\mathcal{E}_y& \mathcal{H} &    \mathcal{P}_x &\mathcal{P}_y &\mathcal{K}_x&\mathcal{K}_y&\mathcal{J} \cr
\mathcal{B}         &  \{ \mathcal{B},\mathcal{B}\}= 0 & 0 & 0 & 0 & 0 & 0 & -\mathcal{E}_y & \mathcal{E}_x & 0\cr
\mathcal{E}_x      &  \{ \mathcal{B},\mathcal{E}_x\}= 0 & 0 & 0 & 0 & 0 & 0 & 0 & c^2\mathcal{B} & \mathcal{E}_y\cr
\mathcal{E}_y     &  \{ \mathcal{B},\mathcal{E}_y\}=0 & 0 & 0 & 0 & 0 & 0 & -c^2\mathcal{B} & 0 & -\mathcal{E}_x\cr
\mathcal{H}        & \{ \mathcal{B},\mathcal{H}\}= 0 & 0 & 0 & 0 & -\mathcal{E}_x & -\mathcal{E}_y & c^2\mathcal{P}_x & c^2\mathcal{P}_y & 0\cr
\mathcal{P}_x     & \{ \mathcal{B},\mathcal{P}_x\}=0 & 0 & 0 & \mathcal{E}_x & 0 & \mathcal{B} & \mathcal{H} & 0 & \mathcal{P}_y\cr
\mathcal{P}_y     &\{ \mathcal{B},\mathcal{P}_y\}= 0 & 0 & 0 & \mathcal{E}_y & -\mathcal{B} & 0 & 0 & \mathcal{H} & -\mathcal{P}_x\cr 
\mathcal{K}_x     & \{ \mathcal{B},\mathcal{K}_x\}=\mathcal{E}_y & 0 &c^2 \mathcal{B} & -c^2\mathcal{P}_x & -\mathcal{H} & 0 & 0 & c^2\mathcal{J} & \mathcal{K}_y\cr
\mathcal{K}_y     &\{ \mathcal{B},\mathcal{K}_y\}=-\mathcal{E}_x & -c^2\mathcal{B} & 0 & -c^2\mathcal{P}_y & 0 & -\mathcal{H} & -c^2\mathcal{J} & 0 & -\mathcal{K}_x\cr
\mathcal{J}         &  \{ \mathcal{B},\mathcal{J}\}=0 & -\mathcal{E}_y & \mathcal{E}_x & 0 & -\mathcal{P}_y & \mathcal{P}_x & -\mathcal{K}_y & \mathcal{K}_x & 0\cr
                  }
                  $$
one quickly notices that they form a family of Poisson algebras parameterised by 
$t$. 
Moreover, the commutation relations are identical for any value of $t$ therefore 
all those algebras are mutually homomorphic.
Therefore, putting $t=0$ to pick a single algebra from that 
  family results in no loss of generality. 
    At the abstract level there is a  $9$ dimensional Lie algebra  $\mathcal{PM}(2+1)$ 
    (a Lie algebra of the Poincar\'e-Maxwell group)
    with the generators 
 $(J_1=B,J_2= E_x,J_3= E_y,J_4 =H,J_5= P_x, J_6=P_y,J_7= K_x, J_8=K_y, J_9=J)$
  characterised by the following, non-vanishing,  brackets
  \begin{equation}
   \begin{tabular}{ l  l}
  $\left[H,K_i \right] = -c^2P_j$,                              & $\left[B,K_i \right] = \varepsilon_{ij} E_j$,  \\
 $\left[P_i,K_j \right] = -\delta_{ij} H$,             &  $\left[H,P_i\right] = E_i$,   \\
 $\left[K_i,K_j \right] = -c^2\varepsilon_{ij} J$,   &  $ \left[E_i,K_j\right] = -c^2\varepsilon_{ij} B$,  \\
$ \left[P_i,J\right] = -\varepsilon_{ij} P_j$,    & $ \left[P_i,P_j\right] = -\varepsilon_{ij} B$,    \\
$ \left[K_i,J\right] = -\varepsilon_{ij} K_j$,    &  $\left[E_i,J\right] = -\varepsilon_{ij} E_j $   \\
\end{tabular} \label{commutators}
  \end{equation}
with $i,j=x,y$.

Note that an algebra of this kind (with $c=1$) was considered in \cite{anyons} where, 
most significantly, the coadjoint action of its Lie group was found.

\section{Coadjoint orbits}\label{coorbit}
In what follows  the coadjoint action of the PM(2+1) group will be presented recapping the results of 
\cite{anyons} by applying the orbit method   
\cite{kirillov1976elements,kirillov1999merits,kirillov2004lectures}.
First of all,  matrices of the adjoint action $m^{ad}_{J_i}$ corresponding to the generators $J_1,\dots, J_9$ are given by
$$
m^{\mathrm{ad}}_{J_i}=
 \begin{bmatrix} 
         c_{i\,1}^1  & \cdots & c_{i\,n}^1 \\
           \vdots    & \ddots & \vdots\\
 c_{i\,1}^n & \cdots & c_{i\,n}^n  
 \end{bmatrix}
$$
where ${c_{ij}}^k$ are the structure constants readily available from  
(\ref{commutators}).
Let us denote by $( \mathcal{B}, \mathcal{E}_x,\mathcal{E}_y, \mathcal{H},    \mathcal{P}_x, \mathcal{P}_y, \mathcal{K}_x, \mathcal{K}_y, \mathcal{J})$
a generic point of the space 
 $\mathcal{PM}^*(2+1)$ 
in the basis dual to the basis { $(B, E_x, E_y, H, P_x, P_y, K_x, K_y, J)$} of 
 $\mathcal{PM}(2+1)$.
 Using the constants of motion found in the previous section as the coordinates 
 will bring about the interpretation of the equations of motion.
The matrix of the coadjoint action corresponding to the group  element 
$$g=
e^{b {B}}
e^{d_x {E_x}}
e^{d_y {E_y}}
e^{\tau {H}}
e^{a_x {P_x}}
e^{a_y {P_y}}
 e^{n_x {K_x}}
e^{n_y {K_y}} 
e^{\phi {J}}
$$
 is given by
 $$M^{coAd}_g=
e^{-\phi m^{ad}_{J}}
 e^{-n_y m^{ad}_{K_y}}
e^{-n_x m^{ad}_{K_x}}
 e^{-a_y m^{ad}_{P_y}}
e^{-a_x m^{ad}_{P_x}}
e^{-\tau m^{ad}_{H}}
e^{-d_y m^{ad}_{E_y}}
e^{-d_x m^{ad}_{E_x}}
e^{-b m^{ad}_{B}}.
$$
 The matrix exponentials of $m^{\mathrm{ad}}_{J_i}$, $i=1,\dots,9$ are presented 
 in the Appendix \ref{appendix}.
 For the explicit form of the coadjoint action of the Poincar\'e-Maxwell group 
 consult \cite{anyons}.
Having established the   coadjoint action of the group $PM(2+1)$ it is time to 
identify its orbits. 
To this end consider the invariants of the coadjoint action, which 
are solutions to the following set of differential equations  
see \cite{Beltrametti196662,doi:10.1063/1.522727,doi:10.1063/1.522992,0305-4470-39-20-009,snobl2014classification}
(note that $AS$ means that the matrix is anti-symmetric)
\begin{equation}
\left[
\begin{array}{ccccccccc}
 0 & 0 & 0 & 0 & 0 & 0 & \mathcal{E}_y & -\mathcal{E}_x & 0 \\
   & 0 & 0 & 0 & 0 & 0 & 0 & -c^2  \mathcal{B}& -\mathcal{E}_y\\
  &   & 0 & 0 & 0 & 0 &  c^2 \mathcal{B}& 0 & \mathcal{E}_x\\
  &   &   & 0 & \mathcal{E}_x & \mathcal{E}_y & -c^2 \mathcal{P}_x & -c^2 \mathcal{P}_y & 0 \\
   &   &   &  & 0 & -\mathcal{B} & -\mathcal{H} & 0 & -\mathcal{P}_y \\
   &   & AS  &  &   & 0 & 0 & -\mathcal{H} & \mathcal{P}_x \\
  &  &     &   & &   & 0 & -c^2 \mathcal{J} & -\mathcal{K}_y \\
  &    &   &  &   &   &   & 0 & \mathcal{K}_x \\
   &   &   &   &   &   &   &  & 0 \\
\end{array}
\right]
\left[
\begin{array}{c}
\frac{\partial}{\partial \mathcal{B}} \\
\frac{\partial}{\partial \mathcal{E}_x} \\
\frac{\partial}{\partial \mathcal{E}_y} \\
\frac{\partial}{\partial \mathcal{H}}\\
\frac{\partial}{\partial \mathcal{P}_x} \\
\frac{\partial}{\partial \mathcal{P}_y}\\
\frac{\partial}{\partial \mathcal{K}_x} \\
\frac{\partial}{\partial \mathcal{K}_y} \\
\frac{\partial }{\partial \mathcal{J}} \\ 
\end{array}
\right]
=
\left[
\begin{array}{c}
0 \\
0 \\
0 \\
0\\
0 \\
0 \\
0 \\
0 \\
0 \\ 
\end{array}
\right].\label{PM_inv_set}
\end{equation}
There are three solutions to (\ref{PM_inv_set}), namely
\begin{align}
    C_0 &=  \vec{\mathcal{E}}^2 - c^2 \vec{\mathcal{B}}^2,\label{PM_inv_C0}\\ 
    C_1 &= \mathcal{H}^2 -c^2 \vec{\mathcal{P}}^2   -  2\left( \vec{\mathcal{K}}\cdot \vec{\mathcal{E}} 
    - c^2 \vec{\mathcal{B}} \cdot \vec{\mathcal{J}}   \right)\label{PM_inv_C1},\\
    C_2 &= \mathcal{H} \vec{\mathcal{B}} + \vec{\mathcal{P}} \times 
    \vec{\mathcal{E}}\label{PM_inv_C2}.
\end{align}
Note that this method of finding the invariants of the coadjoint action does not 
require the knowledge of the explicit form of this action.
Let us consider a generic point  $\xi=\left[ \mathcal{B}, \mathcal{E}_x,\mathcal{E}_y, \mathcal{H},    \mathcal{P}_x, \mathcal{P}_y, \mathcal{K}_x, \mathcal{K}_y, 
\mathcal{J}\right]\in\mathcal{PM}^*(2+1)$,
then the map 
\begin{equation}
  \begin{split}
    C: \mathcal{PM}^*(2+1) &\to \mathbb{R}^3\\
    \xi &\mapsto \left( 
    C_0(\xi),C_1(\xi)\,C_2(\xi)   
    \right)
  \end{split}\label{PM_C_MAP}
\end{equation}
has constant and maximal rank, moreover C is, by construction, constant at each point 
$\xi'=\mathrm{coAd}_q(\xi)$, $g\in PM(2+1)$ belonging to the coadjoint orbit through 
$\xi$.
Quick conclusion is that    the preimage (each of  its compact component, actually) of the point in $\mathbb{R}^3$
is an orbit of the coadjoint action which  will be denoted $\mathcal{O}^{C_0,C_1,C2}$.
Let us attempt to find a parametrisation (equivalently cover it with maps) for 
the orbit $\mathcal{O}^{C_0,C_1,C2}$.
First let us consider a transformation 
\begin{equation}
  \left( \mathcal{B}, \mathcal{E}_x,\mathcal{E}_y, \mathcal{H},    \mathcal{P}_x, \mathcal{P}_y, \mathcal{K}_x, \mathcal{K}_y, \mathcal{J}  \right) 
  \to 
  \left(C_0,C_1,C_2, \mathcal{E}_x,\mathcal{E}_y,     \mathcal{P}_x, \mathcal{P}_y, \mathcal{K}_x, \mathcal{K}_y \right)
\end{equation}
its Jacobian  equals $4\mathcal{B}^3$, therefore as long as $\mathcal{B}\neq 0$ 
the orbit,by the virtue of the implicit function theorem, is locally a graph of some function of $\mathcal{E}_x,\mathcal{E}_y,     \mathcal{P}_x, \mathcal{P}_y, \mathcal{K}_x$ and $\mathcal{K}_y$.
The case $\mathcal{B}=0$ will be commented upon  promptly.
Since $C_0$, $C_1$ and $C_2$ are constant on the orbit,  (\ref{PM_inv_C0}), (\ref{PM_inv_C1}) and (\ref{PM_inv_C2})    
can be used to express $\mathcal{B}$, $\mathcal{H}$ and $\mathcal{J}$ in terms of  
$\mathcal{E}_x$,  $\mathcal{E}_y$, $\mathcal{P}_x$,  $\mathcal{P}_y$, $\mathcal{K}_x$ and  
$\mathcal{K}_y$.
Firstly,
\begin{equation}
  \mathcal{B}(\vec{\mathcal{E}}) =\pm \frac{1}{c} \sqrt{ \vec{\mathcal{E}}^2-C_0}\label{PM_ORB_B}
\end{equation}
which means that for a fixed value of $C_0$ the condition $ \vec{\mathcal{E}}^2 \geq C_0$ 
must hold
for $\mathcal{B}$ ought to be real.
Next, from (\ref{PM_inv_C2}) and (\ref{PM_ORB_B}) 
\begin{equation}
    \mathcal{H}(\vec{\mathcal{E}},\vec{\mathcal{P}}) =   \frac{\vec{\mathcal{E}} \times  \vec{\mathcal{P}}}{ \mathcal{B(\vec{\mathcal{E}})}} 
  + \frac{C_2}{\mathcal{B(\vec{\mathcal{E}})}}\label{PM_ORB_H}
\end{equation}
and finally, keeping in mind (\ref{PM_ORB_B}) and  (\ref{PM_ORB_H}) from (\ref{PM_inv_C1})  one 
has
\begin{equation}
   \mathcal{J}(\vec{\mathcal{E}},\vec{\mathcal{P}},\vec{\mathcal{K}}) =   \frac{
   C_1
   -\mathcal{H}^2(\vec{\mathcal{E}},\vec{\mathcal{P}}) +c^2 \vec{\mathcal{P}}^2
   +2\vec{\mathcal{K}}\cdot \vec{\mathcal{E}} 
   }{2c^2 \mathcal{B}(\vec{\mathcal{E}})}\label{PM_ORB_J}.
\end{equation}

\begin{figure}[h]
  \centering
    \includegraphics[width=0.4\textwidth]{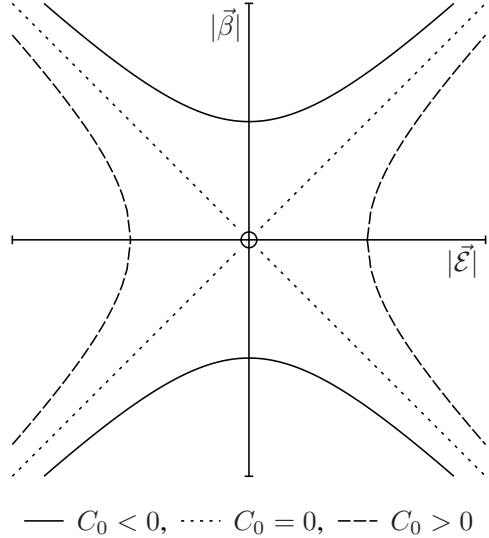}
      \caption{A sketch of the $3$ dimensional cross section of the orbit  
      in the space spanned by $(\mathcal{B},\mathcal{E}_x,\mathcal{E}_x)$
      depicting the cases when $C_0<0$, $C_0=0$ and $C_0>0$.
      }\label{fig1}
\end{figure}

One cannot but notice the ambiguity of (\ref{PM_ORB_B}) that will result in the 
diversity of the orbits. 
When dealing with high dimensional objects the only imagination friendly tool  
one has
at  disposal  consists in taking the cross sections. In that spirit, notice that (\ref{PM_inv_C0})  or
 equivalently (\ref{PM_ORB_B})
defines, in the space spanned by $(\mathcal{B},\mathcal{E}_x,\mathcal{E}_x)$, 
a surface which for $C_0<0$ is hyperboloid of two sheets,  for $C_0=0$ it is a 
conical surface and for $C_0>0$  a hyperboloid of one sheet.
See Figure \ref{fig1}.
It is time to examine all the possibilities one at a time.

First, consider a fixed point $(C_0<0,C_1,C_2)\in \mathbb{R}^3$, its preimage by (\ref{PM_C_MAP}) 
consists of two coadjoint orbits in $\mathcal{PM}^*(2+1)$ denoted 
$\mathcal{O}_{\pm}^{C_0<0,C_1,C_2}$. Note that $\mathcal{B}\neq 0$ in this case.
Each of them can be covered by a single map and in terms of parametrisation 
they are defined by
$$
\varphi: \left(\vec{\mathcal{E}},\vec{\mathcal{P}},\vec{\mathcal{K}}\right)\in 
\mathbb{R}^3
\mapsto 
\left(
  \mathcal{B}(\vec{\mathcal{E}}) ,
\vec{\mathcal{E}},
\mathcal{H}(\vec{\mathcal{E}},\vec{\mathcal{P}}) ,
\vec{\mathcal{P}},
\vec{\mathcal{K}},
 \mathcal{J}(\vec{\mathcal{E}},\vec{\mathcal{P}},\vec{\mathcal{K}}) 
\right)\in \mathcal{PM}^*(2+1)
$$
with $ \mathcal{B}(\vec{\mathcal{E}})$, $\mathcal{H}(\vec{\mathcal{E}},\vec{\mathcal{P}}) $
and $ \mathcal{J}(\vec{\mathcal{E}},\vec{\mathcal{P}},\vec{\mathcal{K}}) $
given by (\ref{PM_ORB_B}), (\ref{PM_ORB_H}) and (\ref{PM_ORB_J}) respectively where $C_0<0$. 

Next, let us examine the preimage of $(C_0=0,C_1,C_2)$. As was said earlier, its three dimensional cross section 
 obtained by fixing $\mathcal{H}$,  $\vec{\mathcal{P}}$,  $\vec{\mathcal{K}}$ and $\mathcal{J}$  
 is a conical surface with a vertex at $(0,0,0)$. Not only does it contain 
 points with $\mathcal{B}=0$, it can not even be given the structure of a smooth 
 manifold.   
The apparent riddle is easily unraveled by realising that
 if $\mathcal{B}=\mathcal{E}_x=\mathcal{E}_y=0$ at any point of the coadjoint 
 orbit it stays constant throughout that orbit. 
 Therefore, this orbit  (the cone is only its cross section) splits into  three parts.  
 The orbit $\mathcal{O}_{\mathcal{B}=0}^{C_0=0,C_1,C_2}$ 
 containing the points for which $\mathcal{B}=\mathcal{E}_x=\mathcal{E}_y=0$ is 
 actually equivalent to the orbit of the Poincar\'e group and  shall not 
be discussed   in depth. 
 Besides that, there are two orbits denoted $\mathcal{O}_{\pm}^{C_0=0,C_1,C_2}$
corresponding to the parts of the cone with positive and negative values of 
$\mathcal{B}$.
Again, each of them can be covered by a single chart and is parametrised by 
$$
\varphi: \left(\vec{\mathcal{E}},\vec{\mathcal{P}},\vec{\mathcal{K}}\right)\in 
\mathbb{R}^3
\mapsto 
\left(
  \mathcal{B}(\vec{\mathcal{E}}) ,
\vec{\mathcal{E}},
\mathcal{H}(\vec{\mathcal{E}},\vec{\mathcal{P}}) ,
\vec{\mathcal{P}},
\vec{\mathcal{K}},
 \mathcal{J}(\vec{\mathcal{E}},\vec{\mathcal{P}},\vec{\mathcal{K}}) 
\right)\in  \mathcal{PM}^*(2+1)
$$
with $ \mathcal{B}(\vec{\mathcal{E}})$, $\mathcal{H}(\vec{\mathcal{E}},\vec{\mathcal{P}}) $
and $ \mathcal{J}(\vec{\mathcal{E}},\vec{\mathcal{P}},\vec{\mathcal{K}}) $
given by (\ref{PM_ORB_B}), (\ref{PM_ORB_H}) and (\ref{PM_ORB_J}) respectively where $C_0=0$.

There is one more type of coadjoint orbits characterised by $C_0>0$. Its cross 
section of constant  $\mathcal{H}$,  $\vec{\mathcal{P}}$,  $\vec{\mathcal{K}}$ and $\mathcal{J}$  
is a hyperboloid of one sheet.
In principle it admits a single parametrisation by means of 
$$
(b,\phi)\mapsto \left( \mathcal{B}=b , \mathcal{E}_x = \sqrt{C_0 - c^2\mathcal{B}^2} \cos{\phi}, \mathcal{E}_y = \sqrt{C_0 - c^2\mathcal{B}^2} \sin{\phi} \right)
$$
with $b\in \mathbb{R}$ and $0 \leq \phi < 2\pi$. 
However, it will be more convenient to assume $\mathcal{B}\neq 0$ 
discarding the case when particle moves in the constant electric field alone.
Doing so, gives two parameterisations  of the regions of the orbit 
disconnected by the plane $\mathcal{B}=0$ which is sufficient for finding 
 the equations of motion since the fields remain constant as the system 
evolves with time. Note that the plane $\mathcal{B}=0$ corresponds to the motion 
of a charged particle in the constant electric field not considered here.
Once again this two regions can be covered by a single chart and are parametrised by 
$$
\varphi: \left(\vec{\mathcal{E}},\vec{\mathcal{P}},\vec{\mathcal{K}}\right)\in 
\mathbb{R}^3
\mapsto 
\left(
  \mathcal{B}(\vec{\mathcal{E}}) ,
\vec{\mathcal{E}},
\mathcal{H}(\vec{\mathcal{E}},\vec{\mathcal{P}}) ,
\vec{\mathcal{P}},
\vec{\mathcal{K}},
 \mathcal{J}(\vec{\mathcal{E}},\vec{\mathcal{P}},\vec{\mathcal{K}}) 
\right)\in  \mathcal{PM}^*(2+1)
$$
with $ \mathcal{B}(\vec{\mathcal{E}})$, $\mathcal{H}(\vec{\mathcal{E}},\vec{\mathcal{P}}) $
and $ \mathcal{J}(\vec{\mathcal{E}},\vec{\mathcal{P}},\vec{\mathcal{K}}) $
given by (\ref{PM_ORB_B}), (\ref{PM_ORB_H}) and (\ref{PM_ORB_J}) respectively where $C_0>0$
 and $\vec{\mathcal{E}}^2> C_0$. 
 It is important to notice that such a $6$ dimensional orbit has no immediate 
 interpretation as a phase space of the system. The problem lies in the fact 
 that besides the conjugate pair $\vec{\mathcal{K}}$ and $\vec{\mathcal{P}}$ the 
 set of coordinates include $\vec{\mathcal{E}}$ which not only do not form a 
 conjugate pair, they are rather artifactual since the fields are constant.

\section{Symplectic structure and dynamics}\label{symplectic}
The parametrisation $\varphi$ can be used to derive the 
equations of motion (briefly recapitulating the results of \cite{anyons} for the most part).
Assuming that $\mathcal{B}\neq 0$ and fixing $C_0$, $C_1$ and $C_2$ one  finds that the 
Poisson structure on the coadjoint orbit, in the map corresponding to the parametrisation $\varphi$ 
is given by
\begin{equation}
\Lambda^{ij}=\left[
\begin{array}{ccccccccc}
 0 & 0  & 0 & 0 & 0 & -c^2  \mathcal{B}\\
 0 & 0  & 0 & 0 &  c^2 \mathcal{B}& 0 \\
 0 & 0 &  0 & -\mathcal{B} & -\mathcal{H} & 0  \\
 0 & 0  & \mathcal{B} & 0 & 0 & -\mathcal{H}  \\
 0 & - c^2 \mathcal{B}   & \mathcal{H} & 0 & 0 & -c^2 \mathcal{J} \\
 c^2\mathcal{B} & 0 &  0 & \mathcal{H} & c^2 \mathcal{J} & 0 \\
\end{array}
\right]\label{PM_orb_poisson}
 \end{equation}
where $\mathcal{B}$, $\mathcal{H}$ and $\mathcal{J}$ are given 
by (\ref{PM_ORB_B}), (\ref{PM_ORB_H}) and (\ref{PM_ORB_J}) respectively. 
The determinant of $\Lambda$ equals $c^8\mathcal{B}^6$ 
which combined with the assumption $\mathcal{B}\neq 0$ 
means that the  Poisson structure (\ref{PM_orb_poisson}) is non-degenerate. 
Quickly inverting the matrix one  finds the corresponding symplectic structure
\begin{equation}
\omega_{ij}=\left[
\begin{array}{ccccccccc}
0 & 
-\frac{\mathcal{ J} }     {{c}^{2}\,\mathcal{B}^{2}} -\frac{  \mathcal{H}^{2}}      {{c}^{4}\,\mathcal{B}^{3}} 
& \frac{\mathcal{H}}{{c}^{2}\,\mathcal{B}^{2}} & 0 & 0 & \frac{1}{{c}^{2}\,\mathcal{B}}\cr 
\frac{\mathcal{ J} }     {{c}^{2}\,\mathcal{B}^{2}} +\frac{  \mathcal{H}^{2}}      {{c}^{4}\,\mathcal{B}^{3}}  & 0 & 0 & -\frac{\mathcal{H}}{{c}^{2}\,\mathcal{B}^{2}} & -\frac{1}{{c}^{2}\,\mathcal{B}} & 0\cr
 -\frac{\mathcal{H}}{{c}^{2}\,\mathcal{B}^{2}} & 0 & 0 & \frac{1}{\mathcal{B}} & 0 & 0\cr 
 0 & \frac{\mathcal{H}}{{c}^{2}\,\mathcal{B}^{2}} & -\frac{1}{\mathcal{B}} & 0 & 0 & 0\cr 
 0 & \frac{1}{{c}^{2}\,\mathcal{B}} & 0 & 0 & 0 & 0\cr 
 -\frac{1}{{c}^{2}\,\mathcal{B}} & 0 & 0 & 0 & 0 & 0
\end{array}\label{symp}
\right].
 \end{equation}
 Finally, the symplectic structure (\ref{symp}) together with the Hamiltonian (\ref{PM_ORB_H}) yields the equations of motion
\begin{equation}
  \dot{\mathcal{E}}_i = 0, \quad
    \dot{\mathcal{P}}_i = - \mathcal{E}_i,\quad      \dot{\mathcal{K}}_i = \mathcal{P}_i . 
    \label{PM_mot1}
\end{equation}
Combining (\ref{PM_mot1}) with (\ref{PM_Pi}) and (\ref{PM_Ki}) yields the familiar equation of motion
\begin{equation}
  \frac{\mathrm{d}}{\mathrm{d} t} \left(  
  \frac{
  m_0 \dot{ \vec{r} }
  }{\sqrt{1    -\frac{   \dot{\vec{r}}^2  }{c^2}     }}
 \right)
 =
 q \left( 
 \dot{\vec{r}} \times \vec{B} + \vec{E}
 \right).\label{PM_mot2}
\end{equation}
Note that the interpretation of (\ref{PM_mot1}), which does not contain the magnetic field, is problematic. 
However, equations (\ref{PM_mot1}) are equivalent to (\ref{PM_mot2}) thus the 
proper interpretation is upheld. Moreover, a clear link between the symplectic dynamics on the coadjoint orbits of the $PM(2+1)$ group 
and the conventional framework of classical mechanics is demonstrated.

 \section{Concluding remarks}\label{conclusions}
Hereby a conclusion arises  that the approach consisting in finding the dynamics on the 
coadjoint orbit of the  Poincar\'e--Maxwell group is equivalent to the standard 
one. See the discussion of the equations of motion on the coadjoint orbit of the  Poincar\'e-Maxwell group
presented in \cite{anyons}.
Although the mathematics involved in finding the coadjoint orbits of the group 
is significantly more complex then the 
standard Hamiltonian mechanics, 
one has to recognise that the Hamiltonian (\ref{PM_ORB_H}) is simpler then the initial one 
(\ref{PM_hamiltonian1}). That simplicity of the Hamiltonian (\ref{PM_ORB_B}) and the fact that the physical 
interpretation remains clear can be seen as an incentive for the further investigations.
One of the possible paths to be taken goes down the road paved by \cite{moyal1} 
where a Stratonovich-Weyl quantisation was performed for the coadjoint orbits of 
the Galilei  group in the free case (also see e.g. \cite{moyal2,moyal3,moyal4}).

\appendix
 \section{Appendix}\label{appendix}
 In this Appendix the matrix exponentials of the 
 matrices of the adjoint action $m^{ad}_{J_i}$ corresponding to the generators $J_1,\dots, J_9$
 are presented.
 {
 
 $$
 e^{b m^{ad}_{B}}=
\begin{bmatrix}
   1 & 0 & 0 & 0 & 0 & 0 & 0 & 0 & 0 \\
 0 & 1 & 0 & 0 & 0 & 0 & 0 & -b & 0 \\
 0 & 0 & 1 & 0 & 0 & 0 & b & 0 & 0 \\
 0 & 0 & 0 & 1 & 0 & 0 & 0 & 0 & 0 \\
 0 & 0 & 0 & 0 & 1 & 0 & 0 & 0 & 0 \\
 0 & 0 & 0 & 0 & 0 & 1 & 0 & 0 & 0 \\
 0 & 0 & 0 & 0 & 0 & 0 & 1 & 0 & 0 \\
 0 & 0 & 0 & 0 & 0 & 0 & 0 & 1 & 0 \\
 0 & 0 & 0 & 0 & 0 & 0 & 0 & 0 & 1 \\
\end{bmatrix}
\,
e^{d_x m^{ad}_{E_x}}=
\begin{bmatrix}
   1 & 0 & 0 & 0 & 0 & 0 & 0 & -c^2 d_x & 0 \\
 0 & 1 & 0 & 0 & 0 & 0 & 0 & 0 & 0 \\
 0 & 0 & 1 & 0 & 0 & 0 & 0 & 0 & -d_x \\
 0 & 0 & 0 & 1 & 0 & 0 & 0 & 0 & 0 \\
 0 & 0 & 0 & 0 & 1 & 0 & 0 & 0 & 0 \\
 0 & 0 & 0 & 0 & 0 & 1 & 0 & 0 & 0 \\
 0 & 0 & 0 & 0 & 0 & 0 & 1 & 0 & 0 \\
 0 & 0 & 0 & 0 & 0 & 0 & 0 & 1 & 0 \\
 0 & 0 & 0 & 0 & 0 & 0 & 0 & 0 & 1 \\
\end{bmatrix}
$$

$$
e^{d_y m^{ad}_{E_y}}=
\begin{bmatrix}
  1 & 0 & 0 & 0 & 0 & 0 & c^2 d_y & 0 & 0 \\
 0 & 1 & 0 & 0 & 0 & 0 & 0 & 0 & d_y \\
 0 & 0 & 1 & 0 & 0 & 0 & 0 & 0 & 0 \\
 0 & 0 & 0 & 1 & 0 & 0 & 0 & 0 & 0 \\
 0 & 0 & 0 & 0 & 1 & 0 & 0 & 0 & 0 \\
 0 & 0 & 0 & 0 & 0 & 1 & 0 & 0 & 0 \\
 0 & 0 & 0 & 0 & 0 & 0 & 1 & 0 & 0 \\
 0 & 0 & 0 & 0 & 0 & 0 & 0 & 1 & 0 \\
 0 & 0 & 0 & 0 & 0 & 0 & 0 & 0 & 1 \\ 
\end{bmatrix}
\,
e^{\tau m^{ad}_{H}}=
\begin{bmatrix}
   1 & 0 & 0 & 0 & 0 & 0 & 0 & 0 & 0 \\
 0 & 1 & 0 & 0 & \tau & 0 & -\frac{c^2 \tau^2}{2}  & 0 & 0 \\
 0 & 0 & 1 & 0 & 0 & \tau & 0 & -\frac{c^2 \tau^2}{2}  & 0 \\
 0 & 0 & 0 & 1 & 0 & 0 & 0 & 0 & 0 \\
 0 & 0 & 0 & 0 & 1 & 0 & -c^2 \tau & 0 & 0 \\
 0 & 0 & 0 & 0 & 0 & 1 & 0 & -c^2 \tau & 0 \\
 0 & 0 & 0 & 0 & 0 & 0 & 1 & 0 & 0 \\
 0 & 0 & 0 & 0 & 0 & 0 & 0 & 1 & 0 \\
 0 & 0 & 0 & 0 & 0 & 0 & 0 & 0 & 1 \\
\end{bmatrix}
$$

$$
e^{a_x m^{ad}_{P_x}}=
\begin{bmatrix}
   1 & 0 & 0 & 0 & 0 & -a_x & 0 & 0 & \frac{a_x^2}{2} \\
 0 & 1 & 0 & -a_x & 0 & 0 & \frac{a_x^2}{2} & 0 & 0 \\
 0 & 0 & 1 & 0 & 0 & 0 & 0 & 0 & 0 \\
 0 & 0 & 0 & 1 & 0 & 0 & -a_x & 0 & 0 \\
 0 & 0 & 0 & 0 & 1 & 0 & 0 & 0 & 0 \\
 0 & 0 & 0 & 0 & 0 & 1 & 0 & 0 & -a_x \\
 0 & 0 & 0 & 0 & 0 & 0 & 1 & 0 & 0 \\
 0 & 0 & 0 & 0 & 0 & 0 & 0 & 1 & 0 \\
 0 & 0 & 0 & 0 & 0 & 0 & 0 & 0 & 1 \\
\end{bmatrix}
 \,
e^{a_y m^{ad}_{P_y}}=
\begin{bmatrix}
   1 & 0 & 0 & 0 & a_y & 0 & 0 & 0 & \frac{a_y^2}{2} \\
 0 & 1 & 0 & 0 & 0 & 0 & 0 & 0 & 0 \\
 0 & 0 & 1 & -a_y & 0 & 0 & 0 & \frac{a_y^2}{2} & 0 \\
 0 & 0 & 0 & 1 & 0 & 0 & 0 & -a_y & 0 \\
 0 & 0 & 0 & 0 & 1 & 0 & 0 & 0 & a_y \\
 0 & 0 & 0 & 0 & 0 & 1 & 0 & 0 & 0 \\
 0 & 0 & 0 & 0 & 0 & 0 & 1 & 0 & 0 \\
 0 & 0 & 0 & 0 & 0 & 0 & 0 & 1 & 0 \\
 0 & 0 & 0 & 0 & 0 & 0 & 0 & 0 & 1 \\
\end{bmatrix}
$$

$$
e^{n_x m^{ad}_{K_x}}=
\begin{bmatrix}
   \cosh {c n_x} & 0 & -c \sinh {cn_x} & 0 & 0 & 0 & 0 & 0 & 0 \\
 0 & 1 & 0 & 0 & 0 & 0 & 0 & 0 & 0 \\
 -\frac{\sinh  {cn_x}}{c} & 0 & \cosh  {cn_x} & 0 & 0 & 0 & 0 & 0 & 0 \\
 0 & 0 & 0 & \cosh {cn_x} & \frac{\sinh  {cn_x}}{c} & 0 & 0 & 0 & 0 \\
 0 & 0 & 0 & c \sinh{cn_x} & \cosh {c n_x} & 0 & 0 & 0 & 0 \\
 0 & 0 & 0 & 0 & 0 & 1 & 0 & 0 & 0 \\
 0 & 0 & 0 & 0 & 0 & 0 & 1 & 0 & 0 \\
 0 & 0 & 0 & 0 & 0 & 0 & 0 & \cosh  {cn_x} & -\frac{\sinh  {c n_x}}{c} \\
 0 & 0 & 0 & 0 & 0 & 0 & 0 & -c \sinh {c n_x} & \cosh {c n_x} \\
\end{bmatrix}
$$

$$
e^{n_y m^{ad}_{K_y}}=
\begin{bmatrix}
   \cosh{c n_y} & c \sinh {c n_y} & 0 & 0 & 0 & 0 & 0 & 0 & 0 \\
 \frac{\sinh {c {n_y}}}{c} & \cosh{c n_y} & 0 & 0 & 0 & 0 & 0 & 0 & 0 \\
 0 & 0 & 1 & 0 & 0 & 0 & 0 & 0 & 0 \\
 0 & 0 & 0 & \cosh {c n_y} & 0 & \frac{\sinh {c n_y}}{c} & 0 & 0 & 0 \\
 0 & 0 & 0 & 0 & 1 & 0 & 0 & 0 & 0 \\
 0 & 0 & 0 & c \sinh {c n_y} & 0 & \cosh{cn_y} & 0 & 0 & 0 \\
 0 & 0 & 0 & 0 & 0 & 0 & \cosh {cn_y} & 0 & \frac{\sinh {c n_y}}{c} \\
 0 & 0 & 0 & 0 & 0 & 0 & 0 & 1 & 0 \\
 0 & 0 & 0 & 0 & 0 & 0 & c \sinh {c n_y} & 0 & \cosh {c n_y}) \\
\end{bmatrix}
$$

$$
e^{\phi m^{ad}_{J}}=
\begin{bmatrix}
   1 & 0 & 0 & 0 & 0 & 0 & 0 & 0 & 0 \\
 0 & \cos{\phi } & -\sin{ \phi } & 0 & 0 & 0 & 0 & 0 & 0 \\
 0 & \sin{\phi}  & \cos{\phi}  & 0 & 0 & 0 & 0 & 0 & 0 \\
 0 & 0 & 0 & 1 & 0 & 0 & 0 & 0 & 0 \\
 0 & 0 & 0 & 0 & \cos {\phi}  & -\sin {\phi}  & 0 & 0 & 0 \\
 0 & 0 & 0 & 0 & \sin{\phi} & \cos {\phi}  & 0 & 0 & 0 \\
 0 & 0 & 0 & 0 & 0 & 0 & \cos{\phi}  & -\sin{\phi}  & 0 \\
 0 & 0 & 0 & 0 & 0 & 0 & \sin{\phi}  & \cos{\phi}  & 0 \\
 0 & 0 & 0 & 0 & 0 & 0 & 0 & 0 & 1 \\
\end{bmatrix}
$$
}

\end{document}